\documentclass[aps,prd,12pt,preprint,tightenlines,superscriptaddress,
amsfonts,amssymb,amsmath,byrevtex,showpacs,nofootinbib]{revtex4}
\usepackage{graphics}
\usepackage{epsfig,subfigure}
\newcounter{mnotecount}[section]
\renewcommand{\themnotecount}{\thesection.\arabic{mnotecount}}
\newcommand{\mnote}[1]
{\protect{\stepcounter{mnotecount}}$^{\mbox{\footnotesize  $
      \bullet$\themnotecount}}$ \marginpar{\raggedright\tiny
    $\!\!\!\!\!\!\,\bullet$\themnotecount: #1} }

\begin{document}
\newcommand{\dR}{\mathbb R}
\newcommand{\dC}{\mathbb C}
\newcommand{\dZ}{\mathbb Z}
\newcommand{\id}{\mathbb I}

\title{Bianchi I model in terms of nonstandard loop quantum cosmology: Classical
dynamics. }

\author{Piotr Dzier\.{z}ak$^\dag$ and W{\l}odzimierz Piechocki$^\ddag$
\\ Theoretical Physics Department, Institute for Nuclear Studies
\\ Ho\.{z}a 69, 00-681 Warsaw, Poland;
\\ $^\dag$pdzi@fuw.edu.pl, $^\ddag$piech@fuw.edu.pl}

\date{\today}

\begin{abstract}
The cosmological singularities of the Bianchi I universe are
analyzed in the setting of loop geometry underlying the loop
quantum cosmology.  We solve the Hamiltonian constraint of the
theory and find the Lie algebra of elementary observables.
Physical compound observables are defined in terms of elementary
ones. Modification of classical theory by holonomy around a loop
removes the singularities. However, our model has a free parameter
that cannot be determined within our method. Testing the model by
the data of observational cosmology may be possible after
quantization of our modified classical theory.
\end{abstract}
 \maketitle

\section{Introduction}

The Bianchi I universe is of primary importance as it  underlies,
to some extent, the Belinskii-Khalatnikov-Lifshitz (BKL) scenario
\cite{BKL,Khalatnikov:2008zt,Garfinkle:2007rv,Erickson:2003zm,Montani:2007vu,MTW},
which is believed to model the Universe in the vicinity of the
cosmological singularity. It has been studied recently
\cite{Bojowald:2003md,Chiou:2006qq,Szulc:2008ar,MartinBenito:2008wx,Ashtekar:2009vc}
within {\it standard} loop quantum cosmology (LQC).

The standard LQC \cite{Ashtekar:2003hd,Bojowald:2006da} means
basically the Dirac method of quantization, which begins with
quantization of the {\it kinematical} phase space followed by
imposition of constraints of the gravitational system in the form
of operators at the quantum level. Finding kernels of these
operators helps to define the {\it physical} Hilbert space. In the
nonstandard LQC  \cite{Dzierzak:2009ip,Malkiewicz:2009qv} one
first solves all the constraints at the classical level to
identify the {\it physical} phase space. Next, one identifies the
algebra of {\it elementary} observables (in the physical phase
space) and finds its representation. Then, {\it compound}
observables are expressed in terms of elementary ones and
quantized. The final goal is finding {\it spectra} of compound
observables which are used to examine the {\it nature} of the
big-bounce phase in the evolution of the Universe.

This paper is devoted to the {\it classical} dynamics of the
Bianchi I model with massless scalar field modified by the loop
geometry, described in the framework of the {\it nonstandard} LQC.
The next paper will address the problem of the quantum dynamics
\cite{PPW}.

In  Sec. II we define the modified classical Hamiltonian. Section
III concerns the final choice of canonical variables. The
classical dynamics is solved in Sec. IV. An algebra of elementary
observables is the subject of Sec. V. Physical observables, that
may be confronted with the cosmological data, are defined  in Sec.
VI. We conclude in the last section. Appendix A presents
derivation of the symplectic form on the constraint hypersurface.
In Appendix B we derive the algebra of elementary observables
corresponding to the Bianchi I model {\it without} the loop
geometry modifications.

\section{Hamiltonian}

The gravitational part of the classical Hamiltonian, $H_g$, in
general relativity is a linear combination of the first-class
constraints, and reads
\begin{equation}\label{hham1}
    H_g:= \int_\Sigma d^3 x (N^i C_i + N^a C_a + N C),
\end{equation}
where $\Sigma$ is the spacelike part of spacetime $\dR \times
\Sigma$, $~(N^i, N^a, N)$ denote Lagrange multipliers, $(C_i, C_a,
C)$ are the Gauss, diffeomorphism and scalar constraint functions.
In our notation  $(a,b = 1,2,3)$ are spatial and $(i,j,k = 1,2,3)$
are internal $SU(2)$ indices. The constraints must satisfy a
specific algebra.

The Bianchi I model with massless scalar field  is described by
the metric:
\begin{equation}\label{bia}
ds^2= -N^2\,dt^2 + \sum_{i=1}^{3} a_i^2(t)\,dx_i^2,
\end{equation}
where
\begin{equation}\label{bbb}
a_i(\tau)=
a_i(0)\,\bigg(\frac{\tau}{\tau_0}\bigg)^{\textrm{k}_i},~~~~d\tau=
N\,dt,~~~~\sum_{i=1}^{3} \textrm{k}_\textrm{i}= 1 = \sum_{i=1}^{3}
\textrm{k}_\textrm{i}^2 + \textrm{k}_{\phi}^2,
\end{equation}
and where $\,\textrm{k}_{\phi}$ describes matter field density
($\textrm{k}_{\phi}= 0$ corresponds to the Kasner model).  For
clear exposition of the {\it singularity} aspects of the Bianchi I
model we recommend \cite{Erickson:2003zm,Chiou:2007mg}.

Having fixed local gauge and diffeomorphism freedom we can rewrite
the gravitational part of the classical Hamiltonian, for the
Bianchi I model with massless scalar field, in the form
\cite{Chiou:2007mg}
\begin{equation}\label{hamG}
H_g = - \gamma^{-2} \int_{\mathcal V} d^3 x ~N
e^{-1}\varepsilon_{ijk}
 E^{aj}E^{bk} F^i_{ab}\, ,
\end{equation}
where  $\gamma$ is the Barbero-Immirzi parameter, $\mathcal
V\subset \Sigma$ is an elementary cell, $\Sigma$ is spacelike
hypersurface,  $N$ denotes the lapse function, $\varepsilon_{ijk}$
is the alternating tensor, $E^a_i $ is a densitized  vector field,
$e:=\sqrt{|\det E|}$, and where $F^i_{ab}$ is the curvature of an
$SU(2)$ connection $A^i_a$.

The resolution of the singularity, obtained within LQC, is based
on rewriting  the curvature $F^k_{ab}$ in terms of holonomies
around loops.   The curvature $F^k_{ab}$   may be determined
 by making use of the formula
\begin{equation}\label{cur}
F^k_{ab}= -2~\lim_{Ar\,\Box_{ij}\,\rightarrow \,0}
Tr\;\Big(\frac{h_{\Box_{ij}}-1}{Ar\,\Box_{ij}}\Big)\;{\tau^k}\;
^o\omega^i_a  \; ^o\omega^j_a ,
\end{equation}
where
\begin{equation}\label{box}
h_{\Box_{ij}} = h^{(\mu_i)}_i h^{(\mu_j)}_j (h^{(\mu_i)}_i)^{-1}
(h^{(\mu_j)}_j)^{-1}
\end{equation}
is the holonomy of the gravitational connection around the square
loop $\Box_{ij}$,  considered over a face of the elementary cell,
each of whose sides has length $\mu_j L_j$ (and $V_o:=L_1 L_2
L_3$) with respect to the flat fiducial metric $^o q_{ab}:=
\delta_{ij}\, ^o \omega^i_a\, ^o \omega^j_a $; the fiducial triad
$^o e^a_k$ and cotriad $^o \omega^k_a$ satisfy $^o \omega^i_a\,^o
e^a_j = \delta^i_j$; $~Ar\,\Box_{ij}$ denotes the area of the
square; and $V_o = \int_{\mathcal V} \sqrt{^o q} d^3 x$ is the
fiducial volume of $\mathcal V$.  In what follows, to simplify
further discussion, we make the assumption $L_1=1=L_2=L_3$ (so
$V_0 =1$), which is natural in the case of choosing  the Bianchi I
model with $T^3$-topology \cite{MartinBenito:2008wx}, instead of
$\dR^3$.

The holonomy in the fundamental, $j=1/2$, representation of
$SU(2)$ reads
\begin{equation}\label{hol}
h^{(\mu_i)}_i  =\cos (\mu_i c_i/2)\;\id + 2\,\sin (\mu_i
c_i/2)\;\tau_i,
\end{equation}
where $\tau_i = -i \sigma_i/2\;$ ($\sigma_i$ are the Pauli spin
matrices). The {\it connection} $A^k_a$ and the density weighted
{\it triad} $E^a_i$ (which occurs in (\ref{identT})) are
determined by the conjugate variables $c$ and $p$:
\begin{equation}
A^i_a = \,c^i\,^o\omega^i_a, ~~~~E^a_i = \,p_i\,^oe^a_i
\end{equation}
where:
\begin{equation}\label{pici}
c_i = \gamma\,\dot{a_i},~~~~ |p_i| = a_j\,a_k
\end{equation}

Making  use of (\ref{hamG}), (\ref{cur})  and the so-called
Thiemann identity
\begin{equation}\label{identT}
\varepsilon_{ijk}\,e^{-1}\,E^{aj}E^{bk} =
\frac{\text{sgn}(p_1p_2p_3)}{2\pi G \gamma
(\mu_1\mu_2\mu_3)^{1/3}}\,\sum_k\,^o\varepsilon^{abc}\,
^o\omega^k_c\,Tr
\Big(h_k^{(\mu_k)}\{(h_k^{(\mu_k)})^{-1},V\}\,\tau_i \Big)
\end{equation}
leads to $H_g$ in the form
\begin{equation}\label{hamR}
    H_g = \lim_{\mu_1,\mu_2,\mu_3\rightarrow \,0}\; H^{(\mu_1\,\mu_2\,\mu_3)}_g ,
\end{equation}
where
\begin{equation}\label{hamL}
H^{(\mu_1\,\mu_2\,\mu_3)}_g = - \frac{\text{sgn}(p_1p_2p_3)}{2\pi
G \gamma^3 \mu_1\mu_2\mu_3} \sum_{ijk}\,N\, \varepsilon^{ijk}\, Tr
\Big(h^{(\mu_i)}_i h^{(\mu_j)}_j (h^{(\mu_i)}_i)^{-1}
(h^{(\mu_j)}_j)^{-1} h_k^{(\mu_k)}\{(h_k^{(\mu_k)})^{-1},V\}\Big),
\end{equation}
and where $V= a_1\,a_2\,a_3$ is the volume of the elementary cell
$\mathcal{V}$.

The total Hamiltonian for Bianchi I universe with a massless
scalar field, $\phi$, reads
\begin{equation}\label{ham}
   H = H_g + H_\phi \approx 0,
\end{equation}
where $H_g$ is defined by (\ref{hamR}). The Hamiltonian of the
scalar field  is known to be: $H_\phi = N\,p^2_\phi
|p|^{-\frac{3}{2}}/2$, where $\phi$ and $p_\phi$ are the
elementary variables satisfying $\{\phi,p_\phi\} = 1$. The
relation $H \approx 0$ defines the {\it physical} phase space of
considered gravitational system with constraints.

Making use of (\ref{hol}) we calculate  (\ref{hamL}) and get the
{\it modified}  total Hamiltonian $H^{(\lambda)}_g$ corresponding
to (\ref{ham}) in the form
\begin{equation}\label{regH}
  H^{(\lambda)}/N= -\frac{1}{8\pi G \gamma^2}\;\frac{\text{sgn}(p_1p_2p_3)}
  {\mu_1\mu_2\mu_3}\bigg[\sin(c_1 \mu_1)\sin(c_2
\mu_2)\,\mu_3\;\textrm{sgn}(p_3)\sqrt{\frac{|p_1p_2|}{|p_3|}} +
\textrm{cyclic}\bigg] + \frac{p_{\phi}^2}{2\,\sqrt{V}}
\end{equation}
where
\begin{equation}\label{re1}
\mu_i:= \sqrt{\frac{1}{|p_i|}}\,\lambda ,
\end{equation}
and where $\lambda$ is a {\it regularization} parameter. Here we
wish to emphasize that (\ref{regH}) presents a  modified {\it
classical} Hamiltonian. It includes {\it no quantum} physics!

In the gauge $N= \sqrt{|p_1\,p_2\,p_3|}$ the Hamiltonian modified
by loop geometry reads
\begin{equation}
H^{(\lambda)}= -\frac{1}{8\pi G
\gamma^2\lambda^2}\;\bigg[|p_1p_2|^{3/2}\sin(c_1 \mu_1)\sin(c_2
\mu_2) + \textrm{cyclic} \bigg] + \frac{p_{\phi}^2}{2} .
\end{equation}

The Poisson bracket is defined to be
\begin{equation}\label{re2}
    \{\cdot,\cdot\}:= 8\pi G\gamma\;\sum_{k=1}^3\bigg[ \frac{\partial \cdot}
    {\partial c_k} \frac{\partial \cdot}{\partial p_k} -
     \frac{\partial \cdot}{\partial p_k} \frac{\partial \cdot}{\partial c_k}\bigg] +
     \frac{\partial \cdot}{\partial \phi} \frac{\partial \cdot}{\partial p_\phi} -
     \frac{\partial \cdot}{\partial p_\phi} \frac{\partial \cdot}{\partial
     \phi} ,
\end{equation}
where $(c_1,c_2,c_3,p_1,p_2,p_3,\phi,p_\phi)$ are canonical
variables.  The dynamics of  $\xi$ reads
\begin{equation}\label{dyn}
    \dot{\xi} := \{\xi,H^{(\lambda)}\},~~~~~~\xi \in
    \{c_1,c_2,c_3,p_1,p_2,p_3,\phi,p_\phi\}.
\end{equation}
The dynamics in the  {\it physical} phase space,
$\mathcal{F}_{phys}^{(\lambda)}$, is defined by solutions to
(\ref{dyn}) satisfying the condition $H^{(\lambda)}\approx 0$. The
solutions of (\ref{dyn}) ignoring the constraint
$H^{(\lambda)}\approx 0$ are in the {\it kinematical} phase space,
$\mathcal{F}_{kin}^{(\lambda)}$.

\section{New canonical  variables}

We use the following canonical variables
\begin{equation}\label{ham1}
\beta_i := \frac{c_i}{\sqrt{|p_i|}},~~~~v_i:=
|p_i|^{3/2}\,\text{sgn}(p_i),
\end{equation}
where $i = 1,2,3$. They satisfy the algebra
\begin{equation}\label{ham2}
\{\beta_i, v_j\}= 12\pi G\gamma\delta_{ij},
\end{equation}
where the Poisson bracket reads
\begin{equation}\label{ham3}
\{\cdot,\cdot\} = 12\pi G\gamma\;\sum_{k=1}^3\bigg[ \frac{\partial
\cdot}{\partial \beta_k} \frac{\partial \cdot}{\partial v_k} -
\frac{\partial \cdot}{\partial v_k} \frac{\partial \cdot}{\partial
\beta_k}\bigg] + \frac{\partial \cdot}{\partial \phi}
\frac{\partial \cdot}{\partial p_\phi} - \frac{\partial
\cdot}{\partial p_\phi} \frac{\partial \cdot}{\partial \phi} .
\end{equation}
The Hamiltonian in the variables (\ref{ham1}) turns out to be
\begin{eqnarray}\nonumber
H^{(\lambda)}= \frac{p_{\phi}^2}{2} - \frac{1}{8\pi
G\gamma^2}\bigg(\frac{\sin(\lambda\beta_1)\sin(\lambda\beta_2)}{\lambda^2}v_1v_2
+\frac{\sin(\lambda\beta_1)\sin(\lambda\beta_3)}{\lambda^2}v_1v_3\\
\label{hamNN}
+\frac{\sin(\lambda\beta_2)\sin(\lambda\beta_3)}{\lambda^2}v_2v_3
\bigg),
\end{eqnarray}
where $\lambda$ parametrizes the holonomy of connection modifying
the Bianchi I model.

\section{Dynamics}

\subsection{Equations of motion}

The Hamilton equations of motion read\footnote{where $i,j,k=1,2,3$
and $i\neq j\neq k$}
\begin{eqnarray}
\dot{\beta_i}&=&
-18\pi G\,\frac{\sin(\lambda\beta_i)}{\lambda}\,(\textrm{O}_\textrm{j}+\textrm{O}_\textrm{k}),\label{beta}\\
\dot{v_i}&=& 18\pi G\,v_i\cos(\lambda\beta_i)\,(\textrm{O}_\textrm{j}+\textrm{O}_\textrm{k}),\label{v}\\
\dot{\phi}&=& p_{\phi},\label{phi}\\
\dot{p_{\phi}}&=& 0,\\
\label{dyn1} H^{(\lambda)}&\approx&0,
\end{eqnarray}
where
\begin{equation}\label{row1}
\textrm{O}_\textrm{i}:= \frac{v_i\,\sin(\lambda\beta_i)}{12\pi G
\gamma\lambda}.
\end{equation}

\subsection{Solution to equations of motion}

Insertion of (\ref{v}) into (\ref{beta}) gives
\begin{eqnarray}
d\beta_i= -\frac{\tan(\lambda\beta_i)}{\lambda}\,\frac{dv_i}{v_i},
\end{eqnarray}
which leads to
\begin{equation}\label{row2}
v_i\,\frac{\sin(\lambda\beta_i)}{\lambda} = \textrm{const}
\end{equation}
Therefore, $\textrm{O}_\textrm{i}$ are constants of motion.

Making use of (\ref{phi}), (\ref{v}) and $\;\cos(\lambda\beta_i)=
\sqrt{1-\sin(\lambda\beta_i)^2}\;$ gives
\begin{equation}\label{eq1}
\int\frac{d v_i}{\sqrt{v_i^2-(12\pi
G\gamma\lambda\,\textrm{O}_\textrm{i})^2}}=18\pi G
\int\frac{(\textrm{O}_\textrm{j}+\textrm{O}_\textrm{k})}{p_{\phi}}\,d\phi.
\end{equation}

Integration of  (\ref{eq1}) leads to
\begin{eqnarray}
\ln\bigg|v_i + \sqrt{v_i^2-(12\pi
G\gamma\lambda\,\textrm{O}_\textrm{i})^2}\bigg|= \frac{18\pi
G}{p_{\phi}}\,(\textrm{O}_\textrm{j}+\textrm{O}_\textrm{k})\,(\phi
- \phi^{0}_i).
\end{eqnarray}
Thus we have
\begin{eqnarray}
2\,|v_i|= \exp\bigg(\frac{18\pi
G}{p_{\phi}}\,(\textrm{O}_\textrm{j}+\textrm{O}_\textrm{k})\,(\phi
- \phi^{0}_i)\bigg) + (12\pi
G\gamma\lambda\textrm{O}_\textrm{i})^2\times\\\nonumber
\times\exp\bigg(-\frac{18\pi
G}{p_{\phi}}\,(\textrm{O}_\textrm{j}+\textrm{O}_\textrm{k})\,(\phi
- \phi^{0}_i)\bigg),
\end{eqnarray}
which may be rewritten as
\begin{equation}\label{eq2}
v_i= 12\pi
G\gamma\lambda\,\,\textrm{O}_\textrm{i}\,\cosh\bigg(\frac{18\pi
G}{p_{\phi}}\,(\textrm{O}_\textrm{j}+\textrm{O}_\textrm{k})\,(\phi
- \phi^{0}_i) - \ln\big|12\pi G\gamma\lambda\textrm{O}_\textrm{i}
\big|\bigg).
\end{equation}

It results from the above solutions that for a nonzero value of
$\lambda$ there is no Big Bang type singularity (for any value of
$\phi$). The Big Bang is replaced by the Big Bounce.  In
\cite{Chiou:2007mg} one considers the so-called {\it planar
collapse}, but we do not consider this issue here as we are mainly
concerned with an {\it initial} type singularity.

Using (\ref{eq2}) it is not difficult to get
\begin{equation}
\sin(\lambda\beta_i)= \frac{1}{\cosh\bigg(\frac{18\pi
G}{p_{\phi}}\,(\textrm{O}_\textrm{j}+\textrm{O}_\textrm{k})\,(\phi
- \phi^{0}_i) - \ln\big|12\pi G\gamma\lambda\textrm{O}_\textrm{i}
\big|\bigg)}\label{rosinus}.
\end{equation}
\\Alternatively, one may solve the equation of motion to get $\beta$.
From  (\ref{phi}) and (\ref{beta}) we obtain
\begin{equation}
\int{\frac{\lambda\,\,d\beta_i}{\sin(\lambda\beta_i)}}= -18\pi
G\,\int{\frac{(\textrm{O}_\textrm{j}+\textrm{O}_\textrm{k})}{
p_{\phi}}\,\,d\phi}\label{calka}.
\end{equation}
The integration of  (\ref{calka}) gives
\begin{equation}
\ln{\bigg|\tan\bigg(\frac{\lambda\beta_i}{2}\bigg)\bigg|}= -18\pi
G\,\frac{(\textrm{O}_\textrm{j}+\textrm{O}_\textrm{k})}{p_{\phi}}\,\phi
+ \textrm{const}.
\end{equation}

Removing the cosmological singularities does not complete our
task. In what follows we consider an algebra of {\it elementary}
observables and physical {\it compound} observables as they define
the background of the nonstandard LQC
\cite{Dzierzak:2009ip,Malkiewicz:2009qv}.

\section{Observables}

A function $F$ defined on the phase space is a Dirac observable if
it is a solution to the equation
\begin{equation}\label{ob1}
\left\{F, H^{(\lambda)}\right\} \approx 0.
\end{equation}
An explicite form of (\ref{ob1}) is given by
\begin{eqnarray}
12\pi G\gamma\sum_{i=1}^{3}\bigg(\frac{\partial F}{\partial
\beta_i}\frac{\partial H^{(\lambda)}}{\partial v_i} -
\frac{\partial F}{\partial v_i}\frac{\partial
H^{(\lambda)}}{\partial \beta_i} \bigg) + \frac{\partial
F}{\partial \phi}p_{\phi} = 0,
\end{eqnarray}
which due to (22) reads
\begin{eqnarray}
18\pi G\,\sum_{i=1}^{3}\bigg[v_i\cos(\lambda\beta_i)\frac{\partial
F}{\partial v_i} -
\frac{\sin(\lambda\beta_i)}{\lambda}\frac{\partial F}{\partial
\beta_i} \bigg]\cdot(\textrm{O}_\textrm{j}+\textrm{O}_\textrm{k})
+ \frac{\partial F}{\partial \phi}p_{\phi}= 0 \label{rowobs}.
\end{eqnarray}

\subsection{Kinematical observables}

One may easily verify that $\textrm{O}_\textrm{i}$ satisfy
(\ref{rowobs}). Instead of solving (\ref{rowobs}) one may use the
constants that occur in (\ref{eq2}) and (\ref{rosinus}). This way
we get
\begin{equation}\label{kin1}
\textrm{A}_\textrm{i} =
\ln{\bigg|\frac{\tan\big(\frac{\lambda\beta_i}{2}\big)}{\frac{\lambda}{2}}\bigg|}
+18\pi
G\,\frac{(\textrm{O}_\textrm{j}+\textrm{O}_\textrm{k})}{p_{\phi}}\,\phi
\end{equation}

The observables (\ref{kin1}) are called {\it kinematical} as they
are not required to satisfy the constraint (\ref{dyn1}).

\subsection{Dynamical observables}

An explicite form of the constraint (\ref{dyn1}) in terms of
$\textrm{O}_\textrm{i}$ is given by
\begin{equation}\label{dy2}
p_{\phi}\,\,\textrm{sgn}(p_{\phi})= 6\sqrt{\pi
G}\,\sqrt{\textrm{O}_\textrm{1}\textrm{O}_\textrm{2} +
\textrm{O}_\textrm{1}\textrm{O}_\textrm{3} +
\textrm{O}_\textrm{2}\textrm{O}_\textrm{3}} .
\end{equation}
It results from (22), (27) and (28) that
$\textrm{O}_\textrm{1}\textrm{O}_\textrm{2} +
\textrm{O}_\textrm{1}\textrm{O}_\textrm{3} +
\textrm{O}_\textrm{2}\textrm{O}_\textrm{3} \geq 0$ so (42) is well
defined. Thus, the {\it dynamical} observables,
$\textrm{A}_\textrm{i}^{\textrm{dyn}}$, corresponding to
(\ref{kin1}) read
\begin{equation}\label{dy3}
\textrm{A}_\textrm{i}^{\textrm{dyn}}=
\ln{\bigg|\frac{\tan\big(\frac{\lambda\beta_i}{2}\big)}{\frac{\lambda}{2}}\bigg|}
+ \frac{3\sqrt{\pi
G}\,\,\textrm{sgn}(p_{\phi})\big(\textrm{O}_\textrm{j}+\textrm{O}_\textrm{k}\big)\,
\phi}{\sqrt{\textrm{O}_\textrm{1}\textrm{O}_\textrm{2} +
\textrm{O}_\textrm{1}\textrm{O}_\textrm{3}  +
\textrm{O}_\textrm{2}\textrm{O}_\textrm{3}}} .
\end{equation}

\subsection{Algebra of observables}

One may verify that  $\textrm{A}_\textrm{i}^{\textrm{dyn}}$
satisfy the following Lie algebra
\begin{eqnarray}
\{\textrm{O}_\textrm{i}, \textrm{O}_\textrm{j}\}&=& 0, \\
\label{aa1} \{\textrm{A}_\textrm{i}^{\textrm{dyn}},
\textrm{O}_\textrm{j}\}&=& \delta_{ij},  \\\label{aa3}
\{\textrm{A}_\textrm{i}^{\textrm{dyn}},\textrm{A}_\textrm{j}^{\textrm{dyn}}\}&=&0.
\end{eqnarray}
In the {\it physical} phase space the Poisson brackets are found
to be (see, Appendix A)
\begin{equation}\label{dyyy}
\{\cdot, \cdot\}_{\textrm{dyn}}:=
\sum_{i=1}^{3}\bigg(\frac{\partial \cdot}{\partial
\textrm{A}_\textrm{i}^{\textrm{dyn}}} \frac{\partial
\cdot}{\partial \textrm{O}_\textrm{i}} - \frac{\partial
\cdot}{\partial \textrm{O}_\textrm{i}} \frac{\partial
\cdot}{\partial \textrm{A}_\textrm{i}^{\textrm{dyn}}} \bigg) ,
\end{equation}
and the algebra reads
\begin{eqnarray}\label{a0}
\{\textrm{O}_\textrm{i}, \textrm{O}_\textrm{j}\}_{\textrm{dyn}}&=& 0, \\
\label{a1} \{\textrm{A}_\textrm{i}^{\textrm{dyn}},
\textrm{O}_\textrm{j}\}_{\textrm{dyn}}&=& \delta_{ij},
 \\\label{a3}
\{\textrm{A}_\textrm{i}^{\textrm{dyn}},\textrm{A}_\textrm{j}
^{\textrm{dyn}}\}_{\textrm{dyn}}&=&0.
\end{eqnarray}

\section{Compound observables}

In what follows we consider the {\it physical} observables which
characterize the singularity aspects of the Bianchi I model. It is
helpful to rewrite (\ref{dy2}) and (\ref{eq2}) in the form
\begin{equation}\label{com1}
p_{\phi}^2= 36\pi G\,\big( \textrm{O}_1\textrm{O}_2 +
\textrm{O}_1\textrm{O}_3 + \textrm{O}_2\textrm{O}_3 \big),
\end{equation}
\begin{equation}\label{com2}
v_i= 12\pi
G\gamma\lambda\,|\,\textrm{O}_\textrm{i}|\,\cosh\bigg(\frac{3\sqrt{\pi
G}\,\,\textrm{sgn}(p_{\phi})\big(\textrm{O}_\textrm{j}+\textrm{O}_\textrm{k}
\big)\,\phi}{\sqrt{\textrm{O}_\textrm{1}\textrm{O}_\textrm{2} +
\textrm{O}_\textrm{1}\textrm{O}_\textrm{3} +
\textrm{O}_\textrm{2}\textrm{O}_\textrm{3}}} +
\ln\bigg(\frac{\lambda}{2}\bigg)-
\textrm{A}_\textrm{i}^{\textrm{dyn}}\bigg).
\end{equation}
The so-called directional energy density \cite{Chiou:2007mg} is
defined to be
\begin{equation}\label{com3}
\rho_{i}(\lambda,\phi):= \frac{p_{\phi}^2}{2\,v_{i}^2}.
\end{equation}

The {\it bounce} in the $i$-th direction occurs when $\rho_{i}$
approaches its maximum \cite{Chiou:2007mg}, which happens at the
minimum of $v_i$ ($p_{\phi}$ is a constant of motion). One may
easily verify that in the case when all three directions coincide,
which corresponds to the Friedmann-Robertson-Walker (FRW) model,
these densities turn into the energy density of the flat FRW with
massless scalar field \cite{Dzierzak:2009ip}.

It is clear that $v_i$ takes minimum  for $\cosh(\cdot)= 1$ so we
have
\begin{equation}\label{den}
v_i^{min}= 12\pi
G\gamma\lambda\,\textrm{O}_\textrm{i},~~~~\rho_{i}^{max}=
\frac{1}{2} \Big(\frac{p_{\phi}}{12\pi
G\gamma\lambda\,\textrm{O}_\textrm{i}}\Big)^2.
\end{equation}
Rewriting $\textrm{O}_i$ and $p_{\phi}$ in terms of $k_i$ and
$k_\phi$ \cite{Chiou:2007mg}
\begin{equation}\label{dd}
\textrm{O}_\textrm{i} = \frac{2}{3}\,\textrm{k}_\textrm{i}K,~~~~
\textrm{p}_{\phi} = \sqrt{8\pi G}\,k_{\phi}K,
\end{equation}
where $K$ is a constant, leads to
\begin{equation}\label{denn}
\rho_{i}^{max}= \frac{1}{16\pi
G\gamma^2\lambda^2}\,\bigg(\frac{\textrm{k}_{\phi}}
{\textrm{k}_i}\bigg)^2 .
\end{equation}
We can determine $\rho_{i}^{max}$ if we know $\lambda$. However,
$\lambda$ is a free parameter of the formalism. Thus, finding the
critical energy densities of matter corresponding to the cosmic
singularities  of the Bianchi I model is an open problem.

One may apply (\ref{denn}) to the Planck scale. Substituting
$\lambda=l_{Pl}$  gives
\begin{equation}\label{de}
\rho_{i}^{max}\simeq
0,35\,\bigg(\frac{k_{\phi}}{k_i}\bigg)^2\,\rho_{Pl},
\end{equation}
which demonstrates that $\rho_{i}^{max}$ may fit the
Planck scale depending on the ratio $k_{\phi}/k_i$.\\

Another important physical  observable is the {\it volume} of the
Universe. From the definitions (\ref{pici}) and (\ref{ham1}) we
get
\begin{equation}
V= a_1 a_2 a_3 = |v_1v_2v_3|^{1/3} .
\end{equation}
It results from (\ref{com2}), (\ref{dd}) and (\ref{bbb}) that the
volume is {\it bounded} from below.

\section{Conclusions}

The {\it modification} of the classical Hamiltonian by using loop
geometry turns classical singularities of the Bianchi I model into
Big Bounces, similarly as in the case of the initial singularity
of the FRW type models \cite{Dzierzak:2009ip}.

Our approach is quite different from the so-called effective or
polymerization method (see, e.g. \cite{Chiou:2007mg}), where the
replacement $\beta \rightarrow \sin(\lambda \beta)/\lambda$ in the
Hamiltonian finishes the procedure of quantization. In our method
this replacement has been done entirely at the {\it classical}
level. Quantization consists in finding a self-adjoint
representation of observables on the {\it physical} phase space
and an examination of the spectra of these observables
\cite{Dzierzak:2009ip,Malkiewicz:2009qv}.

The {\it elementary} observables  constitute a complete set of
constants of motion on the constraint surface. They are used to
parametrize the physical phase space and are ``building blocks''
for the {\it compound} observables like the directional energy
density and the volume operator. So they have deep physical
meaning. Their role becomes even more important at the {\it
quantum} level as they enable finding quantum operators
corresponding to the classical compound observables
\cite{Malkiewicz:2009qv}.

The modification of classical theory by holonomy of connection
around a loop {\it does not} change the Lie algebra of elementary
observables. The algebras of modified (\ref{a0})-(\ref{a3}) and
nonmodified (\ref{h1})-(\ref{h3}) observables are, to some extent,
isomorphic. It is a valuable feature of the modification
procedure.

Our quantum Bianchi I model has a {\it free} parameter $\lambda$
that cannot be determined within the model.  This parameter,
similarly to the case of the FRW  \cite{Dzierzak:2009ip}, is
expected to be fixed by the {\it data} of observational cosmology
(see, e.g. \cite{Mielczarek:2009zw}).

The algebra of elementary observables is defined on the {\it
physical} phase space. The compound observables are thus defined
on the physical phase space too. Thus, their properties may be
confronted with the data of observational cosmology, i.e. {\it
real} world.

 The maximum of the energy densities may fit the Planck scale
which seems to be surprising as the equation used concerns only
the classical level.  But a similar situation occurred in the case
of the FRW model \cite{Dzierzak:2009ip} and the explanation has
come from the examination of the quantum energy density operator.
Its {\it spectrum} turned out to coincide with the classical
counterpart \cite{Malkiewicz:2009qv}.

The present paper gives a background for  {\it quantization} of
the Bianchi I model \cite{PPW}. The quantization is required
despite the fact that the {\it singularity} problem is resolved
already at the classical level due to the modifications based on
the loop geometry via holonomy of connection. It is so because the
{\it spectra} of the quantum observables may be used to get a link
with the data of observational cosmology, similarly as in the case
of the FRW model \cite{Malkiewicz:2009zd, Malkiewicz:2009xz}.
Another reason is that the Big Bounces may occur at any energy
density (being parametrized by a free parameter) so it is
necessary to quantize the model.

Our paper shows that the {\it nonstandard} LQC method, worked out
for the case of the FRW model \cite{Dzierzak:2009ip}, may be
applied to the Bianchi I model. It seems to be applicable to the
isotropic models (e.g. Lema\^{i}tre) as well.

\begin{acknowledgments}
We are grateful to Przemys\l aw Ma\l kiewicz and Wojciech
Zaj\c{a}czkowski for helpful discussions.  WP thanks Haiyun Huang
for interesting correspondence.
\end{acknowledgments}

\appendix

\section{Symplectic form}

The symplectic form on the physical phase space, $\Omega$, may be
obtained from the symplectic form on the kinematical phase space,
$\omega$, by taking into account the constraint (\ref{dy2}).

The symplectic form corresponding to (\ref{ham3}) reads
\begin{eqnarray}
\omega = \frac{1}{12\pi G\gamma}\,\sum_{i=1}^3 d\beta_{i}\wedge
dv_i + d\phi\wedge dp_{\phi}.
\end{eqnarray}
Makin use of the constraint (\ref{dy2}) gives
\begin{eqnarray}
dp_{\phi}= \sum_{i=1}^3 \bigg(\frac{\partial p_{\phi}}{\partial
\beta_{i}}\,d\beta_{i} + \frac{\partial p_{\phi}}{\partial
v_{i}}\,dv_{i}\bigg)= \sum_{i=1}^3 \frac{3\sqrt{\pi
G}\,\,\textrm{sgn}(p_{\phi})\big(\textrm{O}_\textrm{j}+\textrm{O}_\textrm{k}\big)}
{\sqrt{\textrm{O}_\textrm{1}\textrm{O}_\textrm{2} +
\textrm{O}_\textrm{1}\textrm{O}_\textrm{3} +
\textrm{O}_\textrm{2}\textrm{O}_\textrm{3}}}\,d
\textrm{O}_\textrm{i}.
\end{eqnarray}
Thus, we have
\begin{equation}\label{ap1}
d\phi \wedge dp_{\phi}= \sum_{i=1}^3 \frac{3\sqrt{\pi
G}\,\,\textrm{sgn}(p_{\phi})\big(\textrm{O}_\textrm{j}+\textrm{O}_\textrm{k}\big)}
{\sqrt{\textrm{O}_\textrm{1}\textrm{O}_\textrm{2} +
\textrm{O}_\textrm{1}\textrm{O}_\textrm{3} +
\textrm{O}_\textrm{2}\textrm{O}_\textrm{3}}}\,d\phi \wedge d
\textrm{O}_\textrm{i}= \sum_{i=1}^3 \frac{\partial
\textrm{A}_{i}^\textrm{dyn}}{\partial \phi}\,d\phi \wedge d
\textrm{O}_\textrm{i}.
\end{equation}
On the other hand (\ref{row1}) leads to
\begin{eqnarray}
dv_i= \frac{12\pi
G\gamma\lambda}{\sin(\lambda\beta_i)}\,d\textrm{O}_\textrm{i} -
\cot(\lambda\beta_i)\,v_i\,d\beta_i ,
\end{eqnarray}
which gives
\begin{equation}\label{ap2}
d\beta_i \wedge dv_i= \frac{12\pi
G\gamma\lambda}{\sin(\lambda\beta_i)}\,d\beta_i \wedge
d\textrm{O}_\textrm{i}.
\end{equation}
Thus, we have
\begin{eqnarray}
\frac{1}{12\pi G\gamma}\sum_{i=1}^3 d\beta_i \wedge dv_i=
\sum_{i=1}^3 \frac{\lambda}{\sin(\lambda\beta_i)}\,d\beta_i \wedge
d\textrm{O}_i .
\end{eqnarray}
It results from (\ref{ap1}) and (\ref{ap2}) that we have
\begin{equation}\label{aap3}
\Omega = \sum_{i=1}^3
\bigg(\frac{\lambda}{\sin(\lambda\beta_i)}\,d\beta_i +
\frac{\partial \textrm{A}_{\textrm{i}}^\textrm{dyn}}{\partial
\phi}\,d\phi \bigg)\wedge d\textrm{O}_i .
\end{equation}
Now, let us rewrite $d\textrm{A}_{i}^\textrm{dyn}$ as follows
\begin{equation}\label{ap3}
d\textrm{A}_{\textrm{i}}^\textrm{dyn}= \frac{\partial
\textrm{A}_{\textrm{i}}^\textrm{dyn}}{\partial \beta_i}\,d\beta_i
+ \frac{\partial \textrm{A}_{i}^\textrm{dyn}}{\partial v_i}\,dv_i
+ \sum_{j\neq i}\frac{\partial
\textrm{A}_{i}^\textrm{dyn}}{\partial \beta_j}\,d\beta_j +
\sum_{j\neq i}\frac{\partial \textrm{A}_{i}^\textrm{dyn}}{\partial
v_j}\,dv_j + \frac{\partial \textrm{A}_{i}^\textrm{dyn}}{\partial
\phi_i}\,d\phi_i .
\end{equation}
Next, we make summation
\begin{eqnarray}\nonumber
\sum_{i=1}^3 d\textrm{A}_{\textrm{i}}^\textrm{dyn}= \sum_{i=1}^3
\bigg( \frac{\lambda}{\sin(\lambda\beta_i)}\,d\beta_i +
\frac{3\sqrt{\pi
G}\,\textrm{sgn}(p_{\phi})\big(\textrm{O}_\textrm{j}+\textrm{O}_\textrm{k}\big)^2}
{\big(\textrm{O}_1\textrm{O}_2+\textrm{O}_1\textrm{O}_3+
\textrm{O}_2\textrm{O}_3\big)^{3/2}}\,d\textrm{O}_\textrm{i} +
\frac{\partial \textrm{A}_{i}^\textrm{dyn}}{\partial
\phi_i}\,d\phi_i + \\ + \sum_{j\neq i}\frac{\partial
\textrm{A}_{i}^\textrm{dyn}}{\partial \beta_j}\,d\beta_j +
\sum_{j\neq i}\frac{\partial \textrm{A}_{i}^\textrm{dyn}}{\partial
v_j}\,dv_j\bigg).
\end{eqnarray}
Eventually, we calculate the wedge product with
$d\textrm{O}_\textrm{i}$
\begin{eqnarray}
\sum_{i=1}^3 d\textrm{A}_{\textrm{i}}^\textrm{dyn} \wedge
d\textrm{O}_\textrm{i}= \sum_{i=1}^3 \bigg(
\frac{\lambda}{\sin(\lambda\beta_i)}\,d\beta_i + \frac{\partial
\textrm{A}_{\textrm{i}}^\textrm{dyn}}{\partial \phi_i}\,d\phi_i
\bigg) \wedge d \textrm{O}_\textrm{i}.
\end{eqnarray}
Finally, the physical symplectic form $\Omega$ reads
\begin{equation}\label{ap4}
\Omega = \sum_{i=1}^3 d\textrm{A}_{\textrm{i}}^\textrm{dyn} \wedge
d\textrm{O}_\textrm{i}.
\end{equation}
Thus, the physical phase space may be parametrized by the
variables $\textrm{A}_{\textrm{i}}^\textrm{dyn}$ and
$\textrm{O}_\textrm{i}$, and the corresponding Poisson bracket is
given by (\ref{dyyy}).

\section{Nonmodified case}

To find an algebra of elementary observables, for this case, we
introduce the definitions
\begin{eqnarray}\label{obslimit}
\mathcal{O}_i:&=& \lim_{\lambda\rightarrow
\,0}\textrm{O}_\textrm{i}= \frac{v_i\,\beta_i}{12\pi G\gamma},\\
\mathcal{A}_i:&=& \lim_{\lambda\rightarrow
\,0}\textrm{A}_\textrm{i}= \ln{\big|\beta_i\big|} +18\pi
G\,\frac{(\mathcal{O}_j+\mathcal{O}_k)}{p_{\phi}}\,\phi.
\end{eqnarray}
In the limit $\lambda \rightarrow 0$ the Hamiltonian constraint
(\ref{com1}) turns into ``unmodified'' constraint
\begin{equation}\label{hamlimit}
p_{\phi}\,\,\textrm{sgn}(p_{\phi})= 6\sqrt{\pi
G}\,\sqrt{\mathcal{O}_1\mathcal{O}_2 + \mathcal{O}_1\mathcal{O}_3
+ \mathcal{O}_2\mathcal{O}_3} .
\end{equation}
It results from (\ref{obslimit})-(\ref{hamlimit}) that
\begin{equation}\label{Adynlimit}
\mathcal{A}_i^{\textbf{dyn}}= \ln{\big|\beta_i\big|} +
\frac{3\sqrt{\pi
G}\,\,\textrm{sgn}(p_{\phi})(\mathcal{O}_j+\mathcal{O}_k)\,\phi}{\sqrt{\mathcal{O}_1\mathcal{O}_2
+ \mathcal{O}_1\mathcal{O}_3 + \mathcal{O}_2\mathcal{O}_3}}.
\end{equation}
One may verify that
\begin{eqnarray}
\{\mathcal{O}_i, \mathcal{O}_j\}&=& 0, \\
\label{a1lim} \{\mathcal{A}_i^{\textrm{dyn}}, \mathcal{O}_j\}&=&
\delta_{ij},
 \\\label{a3lim}
\{\mathcal{A}_i^{\textrm{dyn}},\mathcal{A}_j^{\textrm{dyn}}\}&=&0,
\end{eqnarray}
and
\begin{equation}\label{symplimit}
\Omega = \sum_{i=1}^3 d\mathcal{A}_i^\textrm{dyn} \wedge
d\mathcal{O}_i .
\end{equation}
Direct calculations lead to
\begin{eqnarray}\label{h1}
\{\mathcal{O}_i, \mathcal{O}_j\}_{\textrm{dyn}}&=& 0, \\
\label{a1dlim} \{\mathcal{A}_i^{\textrm{dyn}},
\mathcal{O}_j\}_{\textrm{dyn}}&=& \delta_{ij},  \\\label{h3}
\{\mathcal{A}_i^{\textrm{dyn}},\mathcal{A}_j
^{\textrm{dyn}}\}_{\textrm{dyn}}&=&0 ,
\end{eqnarray}
which coincides with the algebraic structure of the algebra of the
modified observables (\ref{a0})-(\ref{a3}).

\end{document}